\documentclass[prb, aps, superscriptaddress, amsfonts, twocolumn, showpacs]{revtex4}
\usepackage{graphicx,epstopdf,color}
\usepackage{amsfonts}
\usepackage{amsmath,amssymb,mathrsfs}
\usepackage{bm}
\usepackage[hypertex]{hyperref}

\newcommand{\be}{\begin{equation}}
\newcommand{\ee}{\end{equation}}
\newcommand{\tr}{\text{Tr}}
\newcommand{\lp}{\bigg(}
\newcommand{\rp}{\bigg)}
\newcommand{\lb}{\bigg[}
\newcommand{\rb}{\bigg]}

\renewcommand{\v}[1]{\mathbf{#1}}
\renewcommand{\vr}{\v{r}}
\newcommand{\vk}{\v{k}}
\newcommand{\vg}[1]{\boldsymbol{#1}}
\newcommand{\mean}[1]{\langle#1\rangle}

\newcommand{\ket}[1]{|#1\rangle}
\newcommand{\bra}[1]{\langle#1|}

\begin{document}

\title{Entanglement Entropy of the Two-Dimensional Heisenberg Antiferromagnet}
\author{H.~Francis~Song}
\affiliation{Department of Physics, Yale University, New Haven, CT 06520}
\author{Nicolas Laflorencie}
\affiliation{Laboratoire de Physique des Solides, Universit\'{e} Paris-Sud, UMR-8502 CNRS, 91405 Orsay, France}
\author{Stephan~Rachel}
\affiliation{Department of Physics, Yale University, New Haven, CT 06520}
\author{Karyn~Le~Hur}
\affiliation{Department of Physics, Yale University, New Haven, CT 06520}

\begin{abstract}
We compute the von Neumann and generalized R\'{e}nyi entanglement entropies in the ground-state of the spin-$1/2$ antiferromagnetic Heisenberg model on the square lattice using the modified spin-wave theory for finite lattices. The addition of a staggered magnetic field to regularize the Goldstone modes associated with symmetry-breaking is shown to be essential for obtaining well-behaved values for the entanglement entropy. The von Neumann and R\'{e}nyi entropies obey an area law with additive logarithmic corrections, and are in good quantitative agreement with numerical results from valence bond quantum Monte Carlo and density matrix renormalization group calculations. We also compute the spin fluctuations and observe a multiplicative logarithmic correction to the area law in excellent agreement with quantum Monte Carlo calculations.
\end{abstract}

\date{\today}
\maketitle

\section{Introduction} The scaling of entanglement entropy in the ground-states of quantum many-body Hamiltonians has become a useful tool for characterizing certain universal properties.\cite{Amico08} For the ground-state wavefunction $\ket{\Psi}$ the entanglement entropy is defined as the von Neumman entropy of the reduced density matrix $\hat{\rho}_\Omega=\tr_{\bar{\Omega}}\ket{\Psi}\bra{\Psi}$ of subsystem $\Omega$ obtained by tracing out the degrees of freedom in the remainder of the system $\bar{\Omega}$,
\be
	\mathcal{S}(\hat{\rho}_\Omega)=-\tr(\hat{\rho}_\Omega\ln\hat{\rho}_\Omega).
\ee More generally, the $\alpha$-R\'{e}nyi entropies are defined as
\be
	\mathcal{S}_\alpha(\hat{\rho}_\Omega) = \frac{1}{1-\alpha}\ln[\tr(\hat{\rho}^\alpha_\Omega)] \label{eq:Renyi-def}
\ee and reduce to the von Neumann entropy in the limit $\alpha\rightarrow1$. In one dimension the behavior of the von Neumann and R\'{e}nyi entropies is understood quite well, especially in critical systems which exhibit universal logarithmic scaling with system size,\cite{Calabrese04} but in higher dimensions the von Neumann and R\'{e}nyi entropies are difficult to study. Even numerically, determination of the entanglement entropy is limited to exact diagonalization for small system sizes and the density matrix renormalization group (DMRG) for quasi-one-dimensional systems.\cite{White92,Schollwock05,Kallin09} Recently the R\'{e}nyi entropy $\mathcal{S}_2$ was calculated using quantum Monte Carlo (QMC) in the valence bond basis,\cite{Hastings10} but the von Neumann entropy remained inaccessible in QMC. In dimensions $d>1$ some general results are available for the behavior of the entanglement entropy in free-fermion\cite{Gioev06,Swingle10} and free-boson\cite{Eisert10} theories, and in two dimensions universal terms have also been identified for topologically-ordered phases\cite{Kitaev06,Levin06} and conformal quantum critical points.\cite{Fradkin06,Hsu09} Detailed calculations for specific microscopic Hamiltonians are few, however.

In this work we compute the von Neumann and generalized R\'{e}nyi entropies of the spin-$1/2$ Heisenberg antiferromagnet on the square lattice using the modified spin-wave theory for finite lattices\cite{Takahashi89,Hirsch89,Hirsch89-2} and show that, as expected, the von Neumann and R\'{e}nyi entropies obey an area law \cite{Cramer06} with important additive logarithmic corrections.\cite{Ryu06,Fradkin06,Casini07} For the case of the square lattice with periodic boundary conditions (PBCs) most of the calculations can be done analytically, with the exception of finding the eigenvalues of a matrix. For ladder systems with open boundary conditions (OBCs) we present a simple numerical method that generalizes the procedure for the translationally-invariant case. We performed QMC and DMRG simulations and show that the spin-wave results compare quite favorably with these numerical findings; in addition, there is excellent agreement with recent numerical results.\cite{Kallin09,Hastings10}. We also compute the spin fluctuations and observe a multiplicative logarithmic correction to the area law in excellent agreement with QMC results, further supporting the accuracy of the spin-wave theory.

Although Ref.~\onlinecite{Kallin09} mentions a spin-wave calculation for the entanglement entropy of Heisenberg ladders, this is the first time a proper treatment of the Goldstone modes, which strongly affect the entanglement entropy, has been carried out; previous spin-wave calculations have also missed the additive logarithmic correction.

\section{Modified Spin-Wave Theory} We begin with a careful treatment of the spin-$S$ antiferromagnetic Heisenberg model on a $d$-dimensional hypercubic lattice with sublattices $A$ and $B$, which is described by the Hamiltonian \cite{Manousakis91}
\be
	\hat{H} = \frac{J}{2}\sum_{\vr,\vg{\delta}}\hat{\v{S}}_\vr \cdot \hat{\v{S}}_{\vr+\vg{\delta}} - h\sum_\vr (-1)^\vr\hat{S}^z_\vr\label{eq:H AFHM}
\ee with $J>0$, where $\vg{\delta}$ are the vectors connecting a site on one sublattice with its $z=2d$ nearest neighbors on the other sublattice, and $(-1)^{\vr}=1$ for $\vr\in A$ and $-1$ for $\vr\in B$. We assume PBCs. The staggered field $h>0$ has been included to avoid spurious divergences in finite lattices. Physically, the staggered field regularizes the Goldstone modes associated with the spontaneous breaking of the $\text{O}(3)$ spin-rotational symmetry of Hamiltonian~\eqref{eq:H AFHM} for $h=0$. Indeed, although the spin-wave theory is an expansion around the classical N\'{e}el state in which spins on sublattice $A$ have magnetization $S$ and spins on sublattice $B$ have magnetization $-S$, we will proceed to restore the sublattice symmetry by adjusting $h$ such that $\mean{\hat{S}^z_\vr}=0$. The result of this modified spin-wave theory, treated as a variational wavefunction for Hamiltonian~\eqref{eq:H AFHM} with $h=0$, has been shown to give sensible and remarkably accurate answers for many quantities of interest as compared to quantum Monte Carlo (QMC) calculations.\cite{Hirsch89-2} In this work we show that the spin-wave theory also gives accurate answers for the von Neumann and R\'{e}nyi entanglement entropies, as well as the fluctuations of spin $\hat{S}^z$.

We use the Dyson-Maleev representation to write the spin operators in terms of bosonic creation and annihilation operators $\hat{b}^\dag_\vr,\hat{b}_\vr$ as $\hat{S}^+_\vr = (2S-\hat{n}_\vr)\hat{b}_\vr,\ \hat{S}^-_\vr = \hat{b}^\dag_\vr,\ \hat{S}^z_\vr = S-\hat{n}_\vr$ for $\vr\in A$ and $	\hat{S}^+_\vr = -\hat{b}^\dag_\vr(2S-\hat{n}_\vr),\ \hat{S}^-_\vr = -\hat{b}_\vr,\ \hat{S}^z_\vr = -(S-\hat{n}_\vr)$ for $\vr\in B$, with $\hat{n}_\vr=\hat{b}^\dag_\vr\hat{b}^{}_\vr$. In terms of the bosonic operators we have for $\vr\in A$
\begin{align}
	\hat{\v{S}}_\vr\cdot\hat{\v{S}}_{\vr+\vg{\delta}} &= -S^2 + S(\hat{n}_\vr + \hat{n}_{\vr+\vg{\delta}}
		- \hat{b}_\vr\hat{b}_{\vr+\vg{\delta}} - \hat{b}^\dag_\vr\hat{b}^\dag_{\vr+\vg{\delta}}) \notag\\
		&\qquad + \frac{1}{2}\hat{b}^\dag_\vr(\hat{b}^\dag_{\vr+\vg{\delta}}-\hat{b}_\vr)^2\hat{b}_{\vr+\vg{\delta}},\label{eq:SiSj}
\end{align} so that, after dropping the fourth-order term in Eq.~\eqref{eq:SiSj} (equivalent to making a single-particle density matrix ansatz\cite{Takahashi89}), Hamiltonian~\eqref{eq:H AFHM} becomes the linear spin-wave Hamiltonian
\be
	\hat{H}_\text{LSW} = E_\text{N\'{e}el} + (zJS+h)\sum_\vr\hat{n}_\vr 
		 - \frac{JS}{2}\sum_{\vr,\vg{\delta}}(\hat{b}_\vr\hat{b}_{\vr+\vg{\delta}} + \text{h.c.})\label{eq:H LSW}
\ee where $E_\text{N\'{e}el}/N=-zJS^2/2 - hS$ is the energy of the classical N\'{e}el state. The truncation of the bosonic Hamiltonian has the effect of introducing small mixing between the physical states with $\hat{n}_\vr\leq 2S$ and unphysical states with $\hat{n}_\vr>2S$, which we will see leads to a slight over-estimation of the entanglement entropy due to the larger number of degrees of freedom forming the ground-state.

Although it is conventional to diagonalize Hamiltonian~\eqref{eq:H LSW} treating the two sublattices separately, we can simplify many of the formulas below by treating all bosonic operators as equals. Introducing the Fourier transform $\hat{b}_\vr = (1/\sqrt{N})\sum_\vk e^{-i\vk\cdot\vr}\hat{b}_\vk$ where $N=L^d$ is the total number of sites and $k_\mu=2\pi n_\mu/L$ for $n_\mu=-L/2+1,\ldots,L/2$ in each direction $\mu=1,\ldots,d$, Hamiltonian~\eqref{eq:H LSW} becomes
\be
	\hat{H}_\text{LSW} = E_\text{N\'{e}el} + \frac{zJS}{\eta}\sum_\v{k} \lb \hat{n}_\v{k}
		- \frac{\eta\gamma_\v{k}}{2}(\hat{b}_\v{k} \hat{b}_{-\v{k}} + \text{h.c.}) \rb
\ee with $\hat{n}_\vk=\hat{b}^\dag_\vk\hat{b}^{}_\vk$ and
\be
	\eta = \lp 1+\frac{h}{zJS}\rp^{-1}, \qquad \gamma_\vk = \frac{1}{z}\sum_{\vg{\delta}} e^{i\vk\cdot\vg{\delta}}.
\ee Using the canonical transformation $\hat{\beta}_\v{k} = \cosh\theta_\v{k}\hat{b}_\v{k} - \sinh\theta_\v{k}\hat{b}^\dag_{-\v{k}}$ with $\tanh(2\theta_\v{k}) = \eta\gamma_\v{k}$, Hamiltonian~\eqref{eq:H LSW} is finally written as $\hat{H}_\text{LSW} = E_0 + \sum_\v{k} \omega_\v{k} \hat{\beta}^\dag_\v{k}\hat{\beta}_\v{k}$ where $\omega_\v{k} = (zJS/\eta)\sqrt{1-(\eta\gamma_\v{k})^2}$ and $E_0=E_\text{N\'{e}el}-[zJSN/\eta-\sum_\vk \omega_\vk]/2$ (this is not the variational ground-state energy, however, see Ref.~\onlinecite{Takahashi89}). The single-particle expectation values are
\begin{align}
		\mean{\hat{b}^\dag_\v{r}\hat{b}_{\v{r}'}} = -\frac{1}{2}\delta_{\v{r}\v{r}'} + f(\v{r}-\v{r}'), \ \ \ 
		\mean{\hat{b}_\v{r}\hat{b}_{\v{r}'}} = g(\v{r}-\v{r}'),\label{eq:b b}
\end{align} where
\begin{align}
	f(\v{r}) &= \frac{1}{2N}\sum_\v{k} \cos(\v{k}\cdot\v{r}) \frac{1}{\sqrt{1-(\eta\gamma_\v{k})^2}},\label{eq:fr} \\
	g(\v{r}) &= \frac{1}{2N}\sum_\v{k} \cos(\v{k}\cdot\v{r}) \frac{\eta\gamma_\v{k}}{\sqrt{1-(\eta\gamma_\v{k})^2}}.\label{eq:gr}
\end{align} In particular, $\epsilon = \mean{\hat{n}_\vr} = f(\v{0})-1/2$. We now focus on the square lattice, $d=2$. Typically, we are interested in only the thermodynamic limit $N\rightarrow\infty$ for which the staggered magnetization order parameter at $h=0$ is
\be
	m_z
		= \lim_{h\rightarrow0}\lim_{N\rightarrow\infty}\lb \frac{1}{N}\sum_\vr (-1)^{|\vr|}\mean{\hat{S}^z_\vr}\rb = S - \epsilon_\infty,
\ee where $\epsilon_\infty\simeq 0.197$ after converting Eq.~\eqref{eq:fr} into an integral. For finite $N$, however, the $\vk=(0,0)$ and $\vk=(\pi,\pi)$ Goldstone modes corresponding to the breaking of $\text{O}(3)$ spin-rotational symmetry on each sublattice contribute
\be
	\frac{1}{N}\frac{1}{\sqrt{1-\eta^2}} \simeq \frac{1}{N}\lb\frac{1}{\sqrt{h/2JS}} + O(\sqrt{h/2JS})\rb\label{eq:eps}
\ee to $f(\v{0})$, which shows explicitly how $\epsilon$, and hence $m_z$, diverges for $h\rightarrow0$ unless the limit $N\rightarrow\infty$ is taken first. This is of course the order of limits required for a spontaneously broken symmetry. We can ``repair'' this divergence for finite lattices within the spin-wave approximation by requiring that $m_z=0$, i.e., by adjusting the staggered field $h$ so that $\epsilon=S$:\cite{Takahashi89}
\be
	f(\v{0}) - \frac{1}{2} = S.\label{eq:constraint}
\ee This amounts to restoring the broken sublattice symmetry by hand. As expected, $h\rightarrow0$ as $N\rightarrow\infty$, and indeed for large $N$ we have
\be
	\frac{h}{4JS} \simeq \frac{1}{2m_z^2}\frac{1}{N^2}, \qquad N \gg 1.
\ee Various observables calculated within this formalism have been shown to agree quite well with QMC calculations on finite lattices, without the need to extrapolate to the thermodynamic limit.\cite{Takahashi89,Hirsch89-2} As we will see below, this property is also essential for observing the correct scaling behavior of the entanglement entropy and spin fluctuations with system size.

\section{Spin Fluctuations} Eq.~\eqref{eq:constraint} is the key step to finding well-behaved values for the entanglement entropy for finite system sizes. In particular, it is not sufficient to use small but arbitrary values of $h$ and extrapolate to $h\rightarrow0$. To emphasize this point, we first consider the spin fluctuations in a region $\Omega$,
\be
	\mathcal{F}(\Omega) = \mean{(\hat{S}^z_\Omega - \mean{\hat{S}^z_\Omega})^2}, \qquad \hat{S}^z_\Omega=\sum_{\vr\in \Omega} \hat{S}^z_\vr.
\ee	The fluctuations have been shown to exhibit scaling behavior similar (though not identical) to the entanglement entropy in several systems, including most gapped models and notably in one-dimensional critical systems.\cite{Song10,Hoyos11} In free fermion-like models in any dimension they are related directly to the von Neumann and R\'{e}nyi entropies.\cite{Klich09,Song10-2} While for other models the relation is less direct\cite{Hsu09-2,Cardy10} and in some cases non-existent,\cite{Furukawa11} we can nevertheless use $\mathcal{F}(\Omega)$, which can be computed easily in QMC calculations, as a first check on whether the ground-state obtained in the spin-wave calculation correctly captures non-local correlations. 

The spin fluctuations are computed by summing the spin-spin correlation function as
\be
	\mathcal{F}(\Omega) = \sum_{\vr,\vr'\in \Omega} [\mean{\hat{S}^z_\vr\hat{S}^z_{\vr'}} - \mean{\hat{S}^z_\vr}\mean{\hat{S}^z_{\vr'}}],\label{eq:fluc}
\ee where
\be
	\mean{\hat{S}^z_\vr\hat{S}^z_{\vr'}} - \mean{\hat{S}^z_\vr}\mean{\hat{S}^z_{\vr'}}
	= \frac{1}{3}\epsilon_{\vr\vr'}(\mean{\hat{n}_\vr\hat{n}_{\vr'}} - \mean{\hat{n}_\vr}\mean{\hat{n}_{\vr'}})\label{eq:SzSz corr}
\ee with $\epsilon_{\vr\vr'}=1$ if $\vr,\vr'$ are on the same sublattice and $-1$ otherwise, and, as usual,\cite{Takahashi89} a factor of $1/3$ has been introduced to compensate for the loss of spin-rotational symmetry, $\mean{\hat{\v{S}}_\vr\cdot\hat{\v{S}}_{\vr'}}=\mean{\hat{S}^z_\vr\hat{S}^z_{\vr'}}$ in the present approach. In other words, we consider $\mean{\hat{\v{S}}_\vr\cdot\hat{\v{S}}_{\vr'}}$ as the primary quantity and derive $\mean{\hat{S}^z_\vr\hat{S}^z_{\vr'}}$ from it. Using Eq.~\eqref{eq:b b} and Wick's theorem, we find $\mean{\hat{n}_\vr\hat{n}_{\vr'}} - \mean{\hat{n}_\vr}\mean{\hat{n}_{\vr'}}
	= -\delta_{\vr\vr'}/4 + f(\vr-\vr')^2 + g(\vr-\vr')^2$. Since $g(\vr-\vr')$ vanishes when $\vr,\vr'$ are on the same sublattice while $f(\vr-\vr')$ vanishes when $\vr,\vr'$ are on different sublattices, we obtain
\begin{align}
	&\mean{\hat{S}^z_\vr\hat{S}^z_{\vr'}} - \mean{\hat{S}^z_\vr}\mean{\hat{S}^z_{\vr'}}\notag\\
	&\qquad= \frac{1}{3}\lb -\frac{1}{4}\delta_{\vr\vr'} + f(\vr-\vr')^2 - g(\vr-\vr')^2\rb.\label{eq:SzSz corr3}
\end{align}

\begin{figure}[t]
\hspace{12pt}
\includegraphics*[width=225pt]{fluctuations.eps}\\[-118pt]
\hspace{173pt}\includegraphics*[width=40pt]{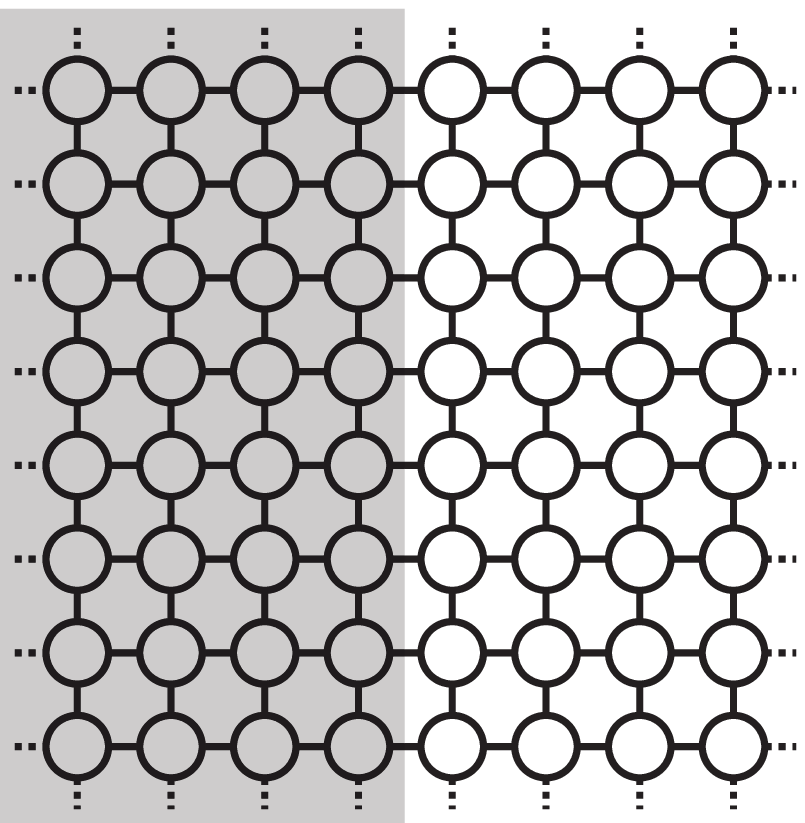}\\[65pt]
\hspace{240pt}
\setlength{\unitlength}{1pt}
	\begin{picture}(0,0)(0,0)
		\put(-110,-7){\makebox(0,0){{\normalsize $L$}}}
		\put(-240,70){\rotatebox{90}{{\normalsize $\mathcal{F}/L$}}}
	\end{picture}\\[5pt]
\caption{(color online). Ground-state spin fluctuations in the spin-$1/2$ Heisenberg antiferromagnet on $L\times L$ square lattices with PBCs. For each $L$ the subsystem is half of the total system, i.e., an $L/2\times L$ region as shown in the inset. Dashed lines show fluctuations computed in the spin-wave approximation for different $L$ at fixed values of the staggered magnetic field. Solid lines show fits to the form $\mathcal{F}(L)=a L\ln L + bL+c$, with $a\simeq0.023,\ b\simeq0.065$ for the spin-wave results and $a\simeq0.023,\ b\simeq0.066$ for the QMC results.}
\label{fig:fluctuations}
\end{figure}
	
The factor of $1/3$ in Eq.~\eqref{eq:SzSz corr} requires some further justification. The spin-wave theory is highly anisotropic, which should not be the case if, for finite lattices at $h=0$, we demand that the spin-rotational symmetry be preserved. For example, we recall that the exact ground-state of Hamiltonian~\eqref{eq:H AFHM} commutes with the total spin operator in the $z$-direction $\hat{S}^z_\text{tot}=\sum_\vr \hat{S}^z_\vr$, which implies that only the diagonal elements of the reduced density matrix for a single spin are non-zero and for spin-$1/2$ take the form
\be
	\rho_\vr = \left(\begin{matrix}
	\frac{1}{2}+\mean{\hat{S}^z_\vr} & 0 \\
	0 & \frac{1}{2}-\mean{\hat{S}^z_\vr}
	\end{matrix}\right).\label{eq:rho single}
\ee With the condition that $\mean{\hat{S}^z_\vr}=0$, we expect $\mean{(\hat{S}^z_\vr)^2}=1/4$, while apparently $\mean{(\hat{S}^z_\vr)^2}=\mean{(\hat{\v{S}}_\vr)^2} = S(S+1)=3/4$. Thus the proper interpretation is that we first compute $\mean{(\hat{\v{S}}_\vr)^2}$ and use the expected relation $\mean{(\hat{S}^x_\vr)^2}=\mean{(\hat{S}^y_\vr)^2}=\mean{(\hat{S}^z_\vr)^2}$ to determine $\mean{(\hat{S}^z_\vr)^2}$, and similarly for more general correlation functions. The factor of $1/3$ is also consistent with the expectation that spin-wave theory slightly underestimates fluctuations relative to QMC as in Fig.~\ref{fig:fluctuations}.

QMC calculations for the Heisenberg antiferromagnet show that $\mathcal{F}(\Omega)$ obeys an area law with a multiplicative logarithmic correction, but for any fixed value of the staggered magnetic field the spin-wave calculations suggest a strict area law as shown in Fig.~\ref{fig:fluctuations}, and moreover, extrapolating to $h\rightarrow0$ does not give the desired result. In contrast, the fluctuations computed using the prescription $\mean{\hat{S}^z_\vr}=0$ show excellent agreement with QMC data, especially for the fit to the form $\mathcal{F}(L)=a L\ln L+bL+c$. We note that a similar multiplicative logarithmic correction was also observed for the valence bond entanglement entropy, which counts the number of singlet bonds crossing the boundary between the subsystem and remainder of the system.\cite{Alet07,Chhajlany07}

%We can understand the multiplicative logarithmic correction as follows. For $N\gg1$ the asymptotic behavior of Eq.~\eqref{eq:SzSz corr3} when %$R=|\v{R}|$, $\v{R}=\vr-\vr'$, is large is
%\begin{align}
%	&\mean{\hat{S}^z_\vr\hat{S}^z_{\vr'}} - \mean{\hat{S}^z_\vr}\mean{\hat{S}^z_{\vr'}}  \notag\\
%	&\qquad \simeq \frac{(-1)^\v{R}}{3}\lp m_z^2 + \frac{\sqrt{2}m_z}{\pi R} + \frac{1}{2\pi^2 R^2} \rp,\label{eq:asymptotic}
%\end{align} where we now consider Eq.~\eqref{eq:SzSz corr3} to hold even for $\v{R}$ that do not connect two actual lattice points, i.e., we %assume Eq.~\eqref{eq:asymptotic} is true in the continuum limit. Thus $(-1)^\v{R}=\cos [\pi R(\cos\theta+\sin\theta)]$ where $\theta$ is the %angle made by $\v{R}$ with the $x$-axis. We note that Eq.~\eqref{eq:asymptotic} also establishes the presence of long-range antiferromagnetic %order in the system in agreement with the traditional approaches.\cite{Takahashi89}

The close agreement between the spin-wave and QMC results gives evidence that the ground-state of Hamiltonian~\eqref{eq:H AFHM} computed in the modified spin-wave theory adequately accounts for non-local correlations as quantified by the spin fluctuations.

\begin{figure}[t]
\hspace{12pt}
\includegraphics*[width=225pt]{hastings.eps}\\[-146pt]
\hspace{10pt}\includegraphics*[width=40pt]{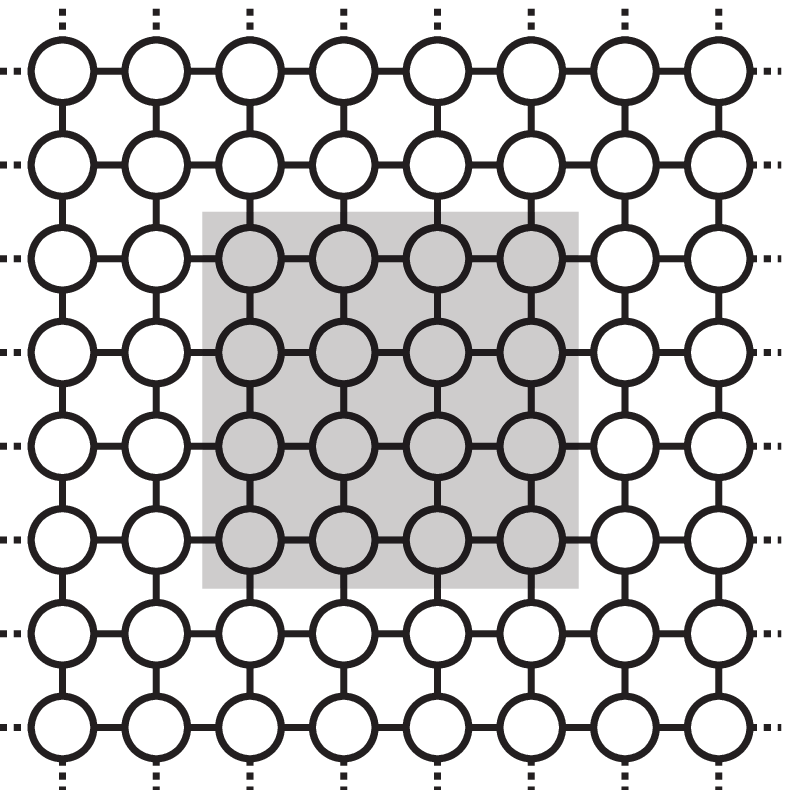}\\[95pt]
\hspace{240pt}
\setlength{\unitlength}{1pt}
	\begin{picture}(0,0)(0,0)
		\put(-111,-6){\makebox(0,0){{\normalsize $\ell$}}}
		\put(-244,73){\rotatebox{90}{{\normalsize $\mathcal{S}_2/\ell$}}}
	\end{picture}\\[5pt]
\caption{(color online). Spin-wave results for the ground-state R\'{e}nyi entropy $\mathcal{S}_2$ in the spin-$1/2$ Heisenberg antiferromagnet on $L\times L$ square lattices with PBCs. For each $L$ the subsystems are $\ell\times\ell$ squares as shown in the inset, cf. Ref.~\onlinecite{Hastings10}.}
\label{fig:hastings}
\end{figure}

\section{Von Neumann and R\'{e}nyi Entanglement Entropies} With the fluctuations showing excellent agreement with QMC results in Fig.~\ref{fig:fluctuations}, we now determine the entanglement entropy within the modified spin-wave theory. Let $X_{\vr\vr'}=\mean{(\hat{b}_\vr + \hat{b}^\dag_\vr)(\hat{b}_{\vr'} + \hat{b}^\dag_{\vr'})}/2$ and $P_{\vr\vr'}=-\mean{(\hat{b}_\vr - \hat{b}^\dag_\vr)(\hat{b}_{\vr'} - \hat{b}^\dag_{\vr'})}/2$, or
\begin{align}
	X_{\vr\vr'} &= f(\vr-\vr') + g(\vr-\vr'),\label{eq:X}\\
	P_{\vr\vr'} &= f(\vr-\vr') - g(\vr-\vr').\label{eq:P}
\end{align} Following Ref.~\onlinecite{Peschel09} (and the equivalent method in Ref.~\onlinecite{Bombelli86}), the entanglement entropy of a region $\Omega$ can be computed from the eigenvalues $\nu_q^2$ of the matrix 
\be
	C_{\vr\vr'} = \sum_{\vr''\in\Omega} X_{\vr\vr''} P_{\vr''\vr'} \label{eq:Crr}
\ee where $\vr,\vr'\in \Omega$. It is important to note that $C_{\vr\vr'}$ is not necessarily symmetric, and a suitable method for finding the eigenvalues must be used. The eigenvalues can be shown to be real and satisfy $\nu_q^2\geq1/4$, however, in accordance with the uncertainty principle. The entanglement entropy is then given by \cite{Peschel09,Casini09}
\begin{align}
	\mathcal{S}(\Omega) &= \sum_q \lb \lp \nu_q+\frac{1}{2}\rp \ln \lp \nu_q + \frac{1}{2}\rp \notag\\
		&\qquad\qquad\qquad- \lp \nu_q-\frac{1}{2}\rp \ln \lp \nu_q - \frac{1}{2}\rp\rb.\label{eq:S nu}
\end{align} In particular, when the subsystem $\Omega$ is a single spin, $C_{\vr\vr'}=f(\v{0})^2$ so that from the constraint~\eqref{eq:constraint} we have $\mathcal{S}(1) = (1+S)\ln(1+S)-S\ln S$. For spin-$1/2$ this gives $\mathcal{S}(1)\simeq 0.955$, which actually over-estimates the exact answer given by the reduced density matrix~\eqref{eq:rho single}, $\ln2\simeq 0.693$. Although particularly extreme for very small subsystem sizes, a slight over-estimation is found to be a general feature of the von Neumann entropy computed in the spin-wave theory, and is most likely due to entanglement arising from the unphysical degrees of freedom that mix with the physical degrees of freedom when the original Hamiltonian is truncated. We can also compute the R\'{e}nyi entropies, which are given by
\be
	\mathcal{S}_\alpha(\Omega) = \frac{1}{\alpha-1}\sum_q\ln [(\nu_q+1/2)^\alpha - (\nu_q-1/2)^\alpha].
\ee 

The second R\'{e}nyi entropy is particularly simple, $\mathcal{S}_2(\Omega) = \sum_q \ln (2\nu_q)$, as is the single-copy entanglement entropy\cite{Eisert05,Peschel05,Orus06} $\mathcal{S}_\infty(\Omega)=\sum_q \ln (\nu_q+1/2)$. For $\mathcal{S}_2$ the single-site R\'{e}nyi entropy from the spin-wave theory is exact for spin-$1/2$: $\mathcal{S}_2(1)=\ln2$, the same as would be from Eq.~\eqref{eq:rho single}; this is special to $\mathcal{S}_2$, however. The R\'{e}nyi entropy $\mathcal{S}_2$ provides another important check on the validity of the spin-wave approximation in computing the entanglement entropy, as it is not clear \emph{a priori} that such a simple theory should properly capture the entanglement content of the ground-state. But $\mathcal{S}_2$ was recently computed in valence bond QMC, and as shown by a comparison of Fig.~\ref{fig:hastings} with Ref.~\onlinecite{Hastings10} the spin-wave results are in remarkably good agreement with numerical data. Spin-wave theory again slightly over-estimates $\mathcal{S}_2$. Fig.~\ref{fig:hastings} also presents $\mathcal{S}_2$ for much larger total system sizes than was possible in Ref.~\onlinecite{Hastings10} to show its behavior essentially without finite-size effects.

\begin{figure}[t]
\hspace{15pt}\includegraphics*[width=225pt]{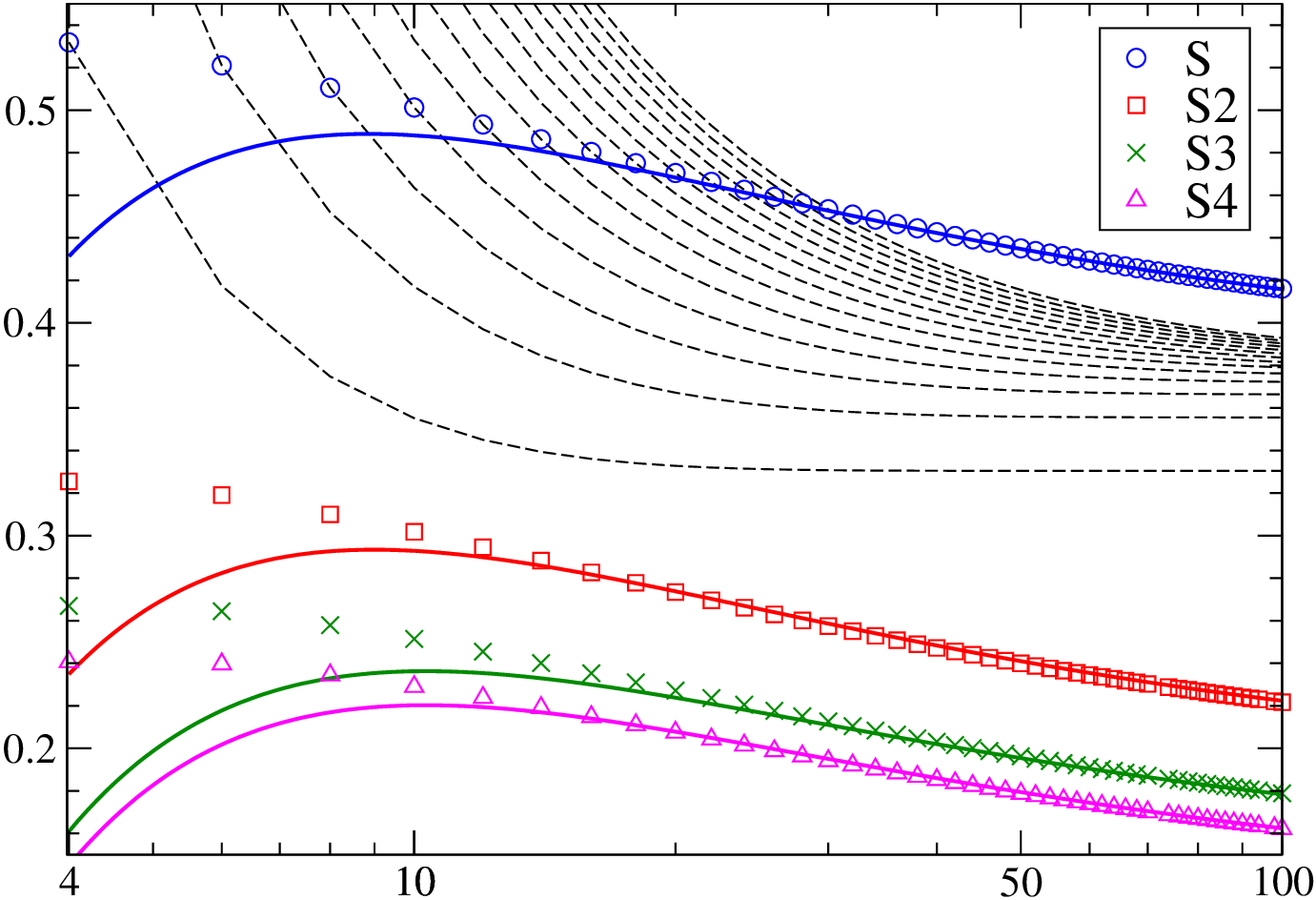}
\setlength{\unitlength}{1pt}
	\begin{picture}(0,0)(0,0)
		\put(-109,-7){\makebox(0,0){{\normalsize $L$}}}
		\put(-246,70){\rotatebox{90}{{\normalsize $\mathcal{S}_\alpha/L$}}}
	\end{picture}\\[5pt]
\caption{(color online). Ground-state von Neumann entanglement entropy $\mathcal{S}$ and R\'{e}nyi entropies $\mathcal{S}_2$, $\mathcal{S}_3$, $\mathcal{S}_4$ in the spin-$1/2$ Heisenberg antiferromagnet on $L\times L$ square lattices with PBCs. The setup is the same as Fig.~\ref{fig:fluctuations}, i.e., subsystems of size $L/2\times L$. Dashed lines show the von Neumann entropy computed in the spin-wave approximation for different $L$ at fixed values of the staggered magnetic field. Solid lines show fits to the form $\mathcal{S}(L)=a L+ b\ln L+c$, cf. Fig.~\ref{fig:renyi-scaling}.}
\label{fig:entropies}
\end{figure}

\begin{figure}[t]
\hspace{15pt}\includegraphics*[width=225pt]{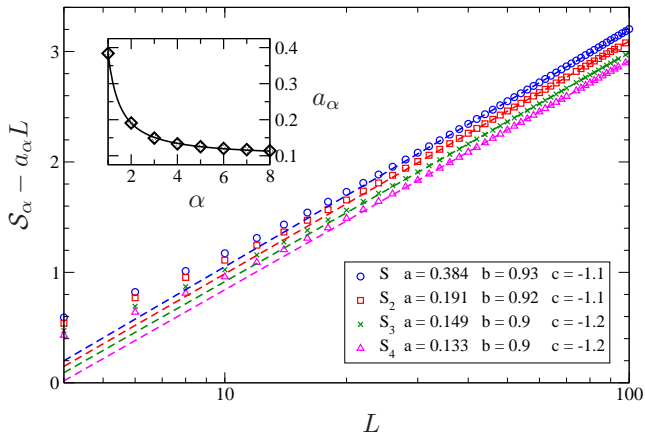}
\setlength{\unitlength}{1pt}
	\begin{picture}(0,0)(0,0)
		\put(-109,-8){\makebox(0,0){{\normalsize $L$}}}
		\put(-246,65){\rotatebox{90}{{\normalsize $\mathcal{S}_\alpha-a_\alpha L$}}}
		\put(-175,76){\makebox(0,0){{\normalsize $\alpha$}}}
		\put(-126,114){\makebox(0,0){{\normalsize $a_\alpha$}}}
	\end{picture}\\[5pt]
\caption{(color online). Extrapolated values for $a,b,c$ in fits of the entanglement entropies of Fig.~\ref{fig:entropies} to the form $\mathcal{S}(L)=a L+b\ln L+c$, indicated by dashed lines. Inset shows the scaling of $a$ with the order of the R\'{e}nyi entropy $\alpha$. Solid line is a fit to $a_\alpha=a_\infty e^{w/\alpha}$ with $a_\infty\simeq0.095$, $w\simeq1.40$.}
\label{fig:renyi-scaling}
\end{figure}

Finally, the spin-wave results for the von Neumann and R\'{e}nyi entropies of the spin-$1/2$ antiferromagnetic Heisenberg model on the square lattice are presented in Fig.~\ref{fig:entropies}. Fits to the form $\mathcal{S}(L)=aL + b\ln L+c$ give $a_1\simeq 0.384$ for $\mathcal{S}_1=\mathcal{S}$ and $a_2\simeq 0.191$ for $\mathcal{S}_2$ after extrapolating to account for the difference in fits for different numbers of points used. The coefficients $b$ and $c$ converge more slowly with the number of points used, but by taking the converged value of $a_\alpha$ and using $\mathcal{S}_\alpha(L)-a_\alpha L$ in a linear fit versus $\ln L$, we find the extrapolated values shown in Fig.~\ref{fig:renyi-scaling}. Interestingly, it appears that the prefactor of the logarithmic term $\ln L$ is essentially the same for all of the R\'{e}nyi entropies including the von Neumann entropy, while the prefactor of the area-law term $aL$ is a function of $\alpha$; numerically, we find that the coefficients $a_\alpha$ for the $\alpha$-R\'{e}nyi entropies very closely follow $a_\alpha= a_\infty e^{w/\alpha}$, inset of Fig.~\ref{fig:renyi-scaling}. For the area-law coefficient of the single-copy entanglement entropy $\mathcal{S}_\infty$, this predicts $a_\infty \simeq 0.095$. According to spin-wave theory, therefore, the leading-order relation $\mathcal{S}_1/\mathcal{S}_\infty\sim 2$ found in Ref.~\onlinecite{Orus06} for critical spin chains does not appear to hold in two dimensions. We note that because of the additive logarithmic correction the entropies do not immediately saturate to a strict area law as is suggested by simply using small values of the staggered field $h$ and extrapolating to $h\rightarrow0$. An additional check on the validity of the spin-wave entanglement entropy is provided below using DMRG data for Heisenberg ladders.

\section{Heisenberg Ladders} For a purely numerical approach to finding the entanglement entropy of spin-wave Hamiltonians that is applicable to arbitrary couplings and boundary conditions, in particular to the Heisenberg ladder geometry of Ref.~\onlinecite{Kallin09}, we consider the Hamiltonian
\be
	\hat{H} = \frac{1}{2}\sum_{ij} J_{ij}\hat{\v{S}}_i\cdot\hat{\v{S}}_j - h\sum_i (-1)^i\hat{S}^z_i,
\ee where $J_{ij}$ is the symmetric matrix describing the coupling between spins. Assuming the expansion about the classical N\'{e}el state remains valid, the Dyson-Maleev transformation leads to the linear spin-wave Hamiltonian
\be
	\hat{H}_B = E_\text{N\'{e}el} + \sum_{ij}\lb A_{ij}\hat{b}^\dag_i\hat{b}_j + \frac{1}{2}B_{ij}(\hat{b}_i\hat{b}_j + \text{h.c.}) \rb\label{eq:LSW general}
\ee with $E_\text{N\'{e}el} = -(S^2/2)\sum_{ij} J_{ij} - hSN,\ A_{ij} = (h + S\sum_k J_{ik}) \delta_{ij},\ B_{ij} = -SJ_{ij}$. We are thus left with the problem of diagonalizing the quadratic boson Hamiltonian in Eq.~\eqref{eq:LSW general}. $A$ and $B$ are real symmetric matrices, and for our purposes we will assume that $A$ is also diagonal. The present approach is well-known,\cite{Bombelli86,Srednicki93,Barthel06,Casini09} but here we describe the steps that lead to a rather straightforward numerical implementation. Switching to first quantization, we use position and momentum variables in the Schr\"{o}dinger representation through the standard relation $\hat{b}=(x+ip)/\sqrt{2}$. Then the Hamiltonian~\eqref{eq:LSW general} becomes (written using vector notation)
\be
	H_B = E_\text{N\'{e}el} - \frac{1}{2}\tr(A) + \frac{1}{2}(p^TGp + x^TVx),\label{eq:Hbosonfirst}
\ee where $G = A-B,\ V = A+B$. We will assume that $V$ and $G$ are both positive-definite, but this must be checked for a given calculation. For the spin-wave Hamiltonian~\eqref{eq:H LSW} with nearest-neighbor coupling and PBCs, for example, the minimum eigenvalue is $h$ for both $G$ and $V$. We then seek the real, symmetric, and positive-definite matrix $W$ such that $WGW=V$, in terms of which Eq.~\eqref{eq:Hbosonfirst} can be written as
\be
	H_B = E_0^B + \frac{1}{2}(p+iWx)^T G(p-iWx).\label{eq:HbosonW}
\ee Since $G$ is positive-definite, if there is a state $\ket{\psi_0}$ that satisfies $(P-iWx)\ket{\psi_0}=0$ then $\ket{\psi_0}$ is the ground-state of Eq.~\eqref{eq:HbosonW} with vacuum energy $E_0^B=E_\text{N\'{e}el}-\tr(A-GW)/2$. In the position basis we have $(\partial_{x_i} + \sum_j W_{ij}x_j )\psi_0(x) = 0$, and it can be easily checked that the normalized solution is $\psi_0(x) = [\det(W/\pi)]^{1/4}\exp(-x^TWx/2)$. The corresponding density matrix $\rho_0(x,x')=\psi_0(x)\psi_0(x')^*$ is 
\be
	\rho_0(x,x') 
		= \lp \det\frac{W}{\pi}\rp^{1/2} \exp\lb -\frac{1}{2}(x^TWx + x'^TWx') \rb.
\ee The matrix $W$ can be found by solving the generalized eigenvalue problem $GVU=UV_D$ in the form
\be
	U^TVU = V_D, \qquad UU^T=G, \label{eq:geneig}
\ee where $V_D$ is a diagonal matrix. Then the inverse of $W$ is given by
\be
	W^{-1} = UV_D^{-1/2}U^T
\ee	and satisfies $W^{-1}VW^{-1}=G$, so that $W$ is found by inverting. If $G$ is the identity as for the Klein-Gordon equation, for example, the conditions~\eqref{eq:geneig} reduce to the usual problem of finding an orthonormal eigenbasis for $V$, $U^{-1}=U^T$, and $W$ is simply the matrix square root of $V$. For the spin-wave Hamiltonian considered here, however, the more general procedure for finding $W$ must be used.

In the Schr\"{o}dinger representation the number operator is given by $\hat{n}_i =(x_i^2 + p_i^2 - 1)/2$, so that
\be
	\epsilon = \frac{1}{N}\sum_i \mean{\hat{n}_i}=\frac{1}{2}\lb \frac{\tr(W^{-1})}{2N} + \frac{\tr(W)}{2N} - 1 \rb.\label{eq:epsnum}
\ee As in Eq.~\eqref{eq:constraint} we impose the constraint $\epsilon=S$, but since in general the Hamiltonian is not translationally invariant the sublattice magnetization is zero only on average and may be non-zero on a given site. Application of Wick's theorem gives
\be
	\mean{\hat{n}_i\hat{n}_j}-\mean{\hat{n}_i}\mean{\hat{n}_j} = -\frac{1}{4}\delta_{ij} + \frac{1}{8}( 
	[W^{-1}]_{ij}^2 + W_{ij}^2),
\ee and the fluctuations can then be calculated using Eqs.~\eqref{eq:fluc}, \eqref{eq:SzSz corr}. Meanwhile, to compute the entanglement entropy of a region $\Omega$, let $\nu_q^2$ be the eigenvalues of $XP$ where
\be
	X=\mean{x_ix_j}=\frac{1}{2}[W^{-1}]_{ij}, \qquad P=\mean{p_ip_j}=\frac{1}{2}W_{ij}
\ee restricted to $i,j\in\Omega$. Then the entanglement entropy is given by Eq.~\eqref{eq:S nu}.

\begin{figure}[t]
\hspace{15pt}\includegraphics*[width=225pt]{kallin.eps}\\[-145pt]
\hspace{130pt}\includegraphics*[width=80pt]{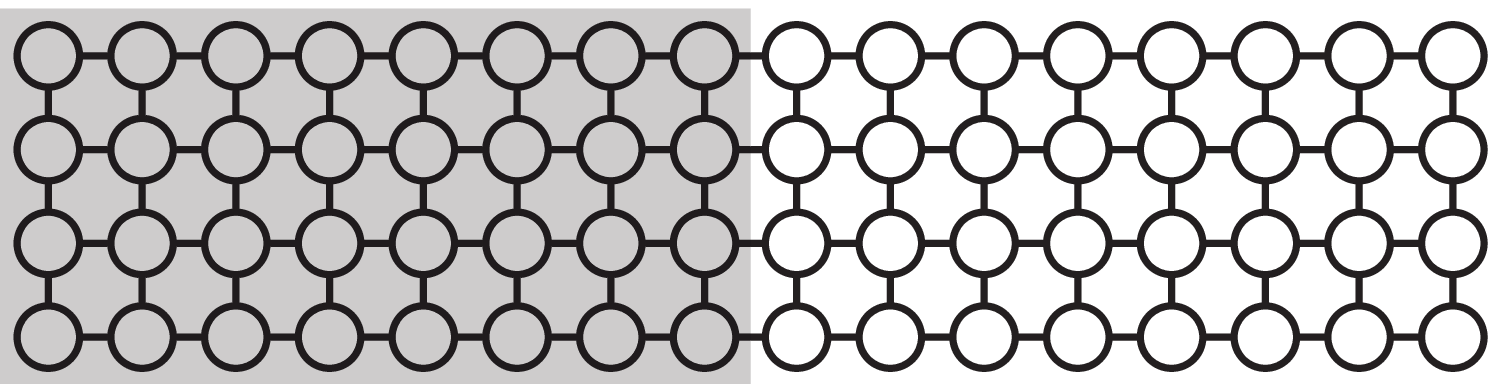}\\[118pt]
\hspace{240pt}
\setlength{\unitlength}{1pt}
	\begin{picture}(0,0)(0,0)
		\put(-107,-5){\makebox(0,0){{\normalsize $N$}}}
		\put(-243,54){\rotatebox{90}{{\normalsize $4\mathcal{F}/N,\ \mathcal{S}_\alpha/N$}}}
	\end{picture}\\[5pt]
\caption{(color online). Ground-state spin fluctuations $\mathcal{F}$ (scaled by 4), von Neumann entanglement entropy $\mathcal{S}$, and R\'{e}nyi entropy $\mathcal{S}_2$ in the spin-$1/2$ antiferromagnetic Heisenberg model on $4N\times N$ ladders with OBCs. As shown in the inset, for each $N$ the subsystem is the left half, i.e., a $2N\times N$ region. Solid (spin-wave) and dashed (DMRG) lines are guides for the eyes only. QMC results for $\mathcal{F}$ are shown as evidence for the convergence of the entropies, especially for $N=7,8$. Cf. Ref.~\onlinecite{Kallin09}.}
\label{fig:kallin}
\end{figure}

Fig.~\ref{fig:kallin} presents DMRG data and spin-wave results for $4N\times N$ Heisenberg ladders with OBCs, showing that the behaviors are similar. The subsystem sizes are $2N\times N$ for each $N$. The fluctuations are again close but slightly underestimated, while the von Neumann and R\'{e}nyi entropies are over-estimated. Although it is difficult to compare numerical fits due to the small number of points for DMRG and even-odd oscillations, the DMRG data appear to approach the spin-wave results for larger $N$ and fits of the spin-wave results to the form $\mathcal{S}=aL+b\ln L+c$ give roughly $a\simeq 0.20$ for the entanglement entropy and $a\simeq 0.10$ for $\mathcal{S}_2$, which are consistent with Fig.~\ref{fig:entropies} after dividing by 2 for OBCs. The spin-wave theory does not perform as well for small $N$, and in particular does not adequately capture the gapped behavior of the even-$N$ ladders,\cite{White94} which is why there are much less oscillations. Again, the presence of the additive logarithmic correction implies that the entropies do not immediately saturate to a strict area law; contrary to previous interpretations \cite{Kallin09} the DMRG data is also consistent with this fact, and QMC data for $\mathcal{F}$ shows that this is not due to lack of convergence.

\section{Conclusions} We have computed the von Neumann and R\'{e}nyi entanglement entropies for the spin-$1/2$ Heisenberg antiferromagnet on the square lattice using the modified spin-wave theory for finite lattices. The results are in good agreement with numerical results, in particular with the R\'{e}nyi entropy $\mathcal{S}_2$ found from valence bond QMC calculations and the entanglement entropy computed in DMRG for small ladders, and confirm the expected area law with important additive logarithmic corrections. We have also computed the spin fluctuations and found excellent agreement with QMC calculations, exhibiting an area law with a multiplicative logarithmic correction. These results support the view that linear spin-wave theory describes very well the ground-state of the spin-$1/2$ Heisenberg antiferromagnet on the square lattice. Interestingly, the essential features of entanglement entropy, namely the area law contribution \emph{and} the important additive logarithmic correction, are captured at this level of approximation.

\section*{Acknowledgements} The work by HFS and KLH was supported by NSF Grant No. DMR-0803200, the Yale Center for Quantum Information Physics (DMR-0653377). SR acknowledges support from the Deutsche Forschungsgemeinschaft under Grant No. RA 1949/1-1. NL acknowledges LPT Toulouse for its hospitality. HFS is very grateful to Priyanka Anand for useful discussions, and SR thanks Peter Schmitteckert for use of his DMRG code.

\bibliography{heisenberg2d}

\begin{thebibliography}{38}
\expandafter\ifx\csname natexlab\endcsname\relax\def\natexlab#1{#1}\fi
\expandafter\ifx\csname bibnamefont\endcsname\relax
  \def\bibnamefont#1{#1}\fi
\expandafter\ifx\csname bibfnamefont\endcsname\relax
  \def\bibfnamefont#1{#1}\fi
\expandafter\ifx\csname citenamefont\endcsname\relax
  \def\citenamefont#1{#1}\fi
\expandafter\ifx\csname url\endcsname\relax
  \def\url#1{\texttt{#1}}\fi
\expandafter\ifx\csname urlprefix\endcsname\relax\def\urlprefix{URL }\fi
\providecommand{\bibinfo}[2]{#2}
\providecommand{\eprint}[2][]{\url{#2}}

\bibitem[{\citenamefont{Amico et~al.}(2008)\citenamefont{Amico, Fazio,
  Osterloh, and Vedral}}]{Amico08}
\bibinfo{author}{\bibfnamefont{L.}~\bibnamefont{Amico}},
  \bibinfo{author}{\bibfnamefont{R.}~\bibnamefont{Fazio}},
  \bibinfo{author}{\bibfnamefont{A.}~\bibnamefont{Osterloh}}, \bibnamefont{and}
  \bibinfo{author}{\bibfnamefont{V.}~\bibnamefont{Vedral}},
  \bibinfo{journal}{Rev. Mod. Phys.} \textbf{\bibinfo{volume}{80}},
  \bibinfo{pages}{517} (\bibinfo{year}{2008}).

\bibitem[{\citenamefont{Calabrese and Cardy}(2004)}]{Calabrese04}
\bibinfo{author}{\bibfnamefont{P.}~\bibnamefont{Calabrese}} \bibnamefont{and}
  \bibinfo{author}{\bibfnamefont{J.}~\bibnamefont{Cardy}}, \bibinfo{journal}{J.
  Stat. Mech.} \textbf{\bibinfo{volume}{(2004)}}, \bibinfo{pages}{{P06002}}
  (\bibinfo{year}{2004}).

\bibitem[{\citenamefont{White}(1992)}]{White92}
\bibinfo{author}{\bibfnamefont{S.~R.} \bibnamefont{White}},
  \bibinfo{journal}{Phys. Rev. Lett.} \textbf{\bibinfo{volume}{69}},
  \bibinfo{pages}{2863} (\bibinfo{year}{1992}).

\bibitem[{\citenamefont{Schollw{\"{o}}ck}(2005)}]{Schollwock05}
\bibinfo{author}{\bibfnamefont{U.}~\bibnamefont{Schollw{\"{o}}ck}},
  \bibinfo{journal}{Rev. Mod. Phys.} \textbf{\bibinfo{volume}{77}},
  \bibinfo{pages}{259} (\bibinfo{year}{2005}).

\bibitem[{\citenamefont{Kallin et~al.}(2009)\citenamefont{Kallin,
  Gonz{\'{a}}lez, Hastings, and Melko}}]{Kallin09}
\bibinfo{author}{\bibfnamefont{A.~B.} \bibnamefont{Kallin}},
  \bibinfo{author}{\bibfnamefont{I.}~\bibnamefont{Gonz{\'{a}}lez}},
  \bibinfo{author}{\bibfnamefont{M.~B.} \bibnamefont{Hastings}},
  \bibnamefont{and} \bibinfo{author}{\bibfnamefont{R.~G.} \bibnamefont{Melko}},
  \bibinfo{journal}{Phys. Rev. Lett.} \textbf{\bibinfo{volume}{103}},
  \bibinfo{pages}{117203} (\bibinfo{year}{2009}).

\bibitem[{\citenamefont{Hastings et~al.}(2010)\citenamefont{Hastings,
  Gonz{\'{a}}lez, Kallin, and Melko}}]{Hastings10}
\bibinfo{author}{\bibfnamefont{M.~B.} \bibnamefont{Hastings}},
  \bibinfo{author}{\bibfnamefont{I.}~\bibnamefont{Gonz{\'{a}}lez}},
  \bibinfo{author}{\bibfnamefont{A.~B.} \bibnamefont{Kallin}},
  \bibnamefont{and} \bibinfo{author}{\bibfnamefont{R.~G.} \bibnamefont{Melko}},
  \bibinfo{journal}{Phys. Rev. Lett.} \textbf{\bibinfo{volume}{104}},
  \bibinfo{pages}{157201} (\bibinfo{year}{2010}).

\bibitem[{\citenamefont{Gioev and Klich}(2006)}]{Gioev06}
\bibinfo{author}{\bibfnamefont{D.}~\bibnamefont{Gioev}} \bibnamefont{and}
  \bibinfo{author}{\bibfnamefont{I.}~\bibnamefont{Klich}},
  \bibinfo{journal}{Phys. Rev. Lett.} \textbf{\bibinfo{volume}{96}},
  \bibinfo{pages}{100503} (\bibinfo{year}{2006}).

\bibitem[{\citenamefont{Swingle}(2010)}]{Swingle10}
\bibinfo{author}{\bibfnamefont{B.}~\bibnamefont{Swingle}},
  \bibinfo{journal}{Phys. Rev. Lett.} \textbf{\bibinfo{volume}{105}},
  \bibinfo{pages}{050502} (\bibinfo{year}{2010}).

\bibitem[{\citenamefont{Eisert et~al.}(2010)\citenamefont{Eisert, Cramer, and
  Plenio}}]{Eisert10}
\bibinfo{author}{\bibfnamefont{J.}~\bibnamefont{Eisert}},
  \bibinfo{author}{\bibfnamefont{M.}~\bibnamefont{Cramer}}, \bibnamefont{and}
  \bibinfo{author}{\bibfnamefont{M.~B.} \bibnamefont{Plenio}},
  \bibinfo{journal}{Rev. Mod. Phys.} \textbf{\bibinfo{volume}{82}},
  \bibinfo{pages}{277} (\bibinfo{year}{2010}).

\bibitem[{\citenamefont{Kitaev and Preskill}(2006)}]{Kitaev06}
\bibinfo{author}{\bibfnamefont{A.}~\bibnamefont{Kitaev}} \bibnamefont{and}
  \bibinfo{author}{\bibfnamefont{J.}~\bibnamefont{Preskill}},
  \bibinfo{journal}{Phys. Rev. Lett.} \textbf{\bibinfo{volume}{96}},
  \bibinfo{pages}{110404} (\bibinfo{year}{2006}).

\bibitem[{\citenamefont{Levin and Wen}(2006)}]{Levin06}
\bibinfo{author}{\bibfnamefont{M.}~\bibnamefont{Levin}} \bibnamefont{and}
  \bibinfo{author}{\bibfnamefont{X.-G.} \bibnamefont{Wen}},
  \bibinfo{journal}{Phys. Rev. Lett.} \textbf{\bibinfo{volume}{96}},
  \bibinfo{pages}{110405} (\bibinfo{year}{2006}).

\bibitem[{\citenamefont{Fradkin and Moore}(2006)}]{Fradkin06}
\bibinfo{author}{\bibfnamefont{E.}~\bibnamefont{Fradkin}} \bibnamefont{and}
  \bibinfo{author}{\bibfnamefont{J.~E.} \bibnamefont{Moore}},
  \bibinfo{journal}{Phys. Rev. Lett.} \textbf{\bibinfo{volume}{97}},
  \bibinfo{pages}{050404} (\bibinfo{year}{2006}).

\bibitem[{\citenamefont{Hsu et~al.}(2009{\natexlab{a}})\citenamefont{Hsu,
  Mulligan, Fradkin, and Kim}}]{Hsu09}
\bibinfo{author}{\bibfnamefont{B.}~\bibnamefont{Hsu}},
  \bibinfo{author}{\bibfnamefont{M.}~\bibnamefont{Mulligan}},
  \bibinfo{author}{\bibfnamefont{E.}~\bibnamefont{Fradkin}}, \bibnamefont{and}
  \bibinfo{author}{\bibfnamefont{E.-A.} \bibnamefont{Kim}},
  \bibinfo{journal}{Phys. Rev. B} \textbf{\bibinfo{volume}{79}},
  \bibinfo{pages}{115421} (\bibinfo{year}{2009}{\natexlab{a}}).

\bibitem[{\citenamefont{Takahashi}(1989)}]{Takahashi89}
\bibinfo{author}{\bibfnamefont{M.}~\bibnamefont{Takahashi}},
  \bibinfo{journal}{Phys. Rev. B} \textbf{\bibinfo{volume}{40}},
  \bibinfo{pages}{2494} (\bibinfo{year}{1989}).

\bibitem[{\citenamefont{Hirsch and Tang}(1989{\natexlab{a}})}]{Hirsch89}
\bibinfo{author}{\bibfnamefont{J.~E.} \bibnamefont{Hirsch}} \bibnamefont{and}
  \bibinfo{author}{\bibfnamefont{S.}~\bibnamefont{Tang}},
  \bibinfo{journal}{Phys. Rev. B} \textbf{\bibinfo{volume}{39}},
  \bibinfo{pages}{2850} (\bibinfo{year}{1989}{\natexlab{a}}).

\bibitem[{\citenamefont{Hirsch and Tang}(1989{\natexlab{b}})}]{Hirsch89-2}
\bibinfo{author}{\bibfnamefont{J.~E.} \bibnamefont{Hirsch}} \bibnamefont{and}
  \bibinfo{author}{\bibfnamefont{S.}~\bibnamefont{Tang}},
  \bibinfo{journal}{Phys. Rev. B} \textbf{\bibinfo{volume}{40}},
  \bibinfo{pages}{4769} (\bibinfo{year}{1989}{\natexlab{b}}).

\bibitem[{\citenamefont{Cramer et~al.}(2006)\citenamefont{Cramer, Eisert,
  Plenio, and Drei{\ss}ig}}]{Cramer06}
\bibinfo{author}{\bibfnamefont{M.}~\bibnamefont{Cramer}},
  \bibinfo{author}{\bibfnamefont{J.}~\bibnamefont{Eisert}},
  \bibinfo{author}{\bibfnamefont{M.~B.} \bibnamefont{Plenio}},
  \bibnamefont{and}
  \bibinfo{author}{\bibfnamefont{J.}~\bibnamefont{Drei{\ss}ig}},
  \bibinfo{journal}{Phys. Rev. A} \textbf{\bibinfo{volume}{73}},
  \bibinfo{pages}{012309} (\bibinfo{year}{2006}).

\bibitem[{\citenamefont{Ryu and Takayanagi}(2006)}]{Ryu06}
\bibinfo{author}{\bibfnamefont{S.}~\bibnamefont{Ryu}} \bibnamefont{and}
  \bibinfo{author}{\bibfnamefont{T.}~\bibnamefont{Takayanagi}},
  \bibinfo{journal}{Phys. Rev. Lett.} \textbf{\bibinfo{volume}{96}},
  \bibinfo{pages}{181602} (\bibinfo{year}{2006}).

\bibitem[{\citenamefont{Casini and Huerta}(2007)}]{Casini07}
\bibinfo{author}{\bibfnamefont{H.}~\bibnamefont{Casini}} \bibnamefont{and}
  \bibinfo{author}{\bibfnamefont{M.}~\bibnamefont{Huerta}},
  \bibinfo{journal}{Nucl. Phys. B} \textbf{\bibinfo{volume}{764}},
  \bibinfo{pages}{183} (\bibinfo{year}{2007}).

\bibitem[{\citenamefont{Manousakis}(1991)}]{Manousakis91}
\bibinfo{author}{\bibfnamefont{E.}~\bibnamefont{Manousakis}},
  \bibinfo{journal}{Rev. Mod. Phys.} \textbf{\bibinfo{volume}{63}},
  \bibinfo{pages}{1} (\bibinfo{year}{1991}).

\bibitem[{\citenamefont{Song et~al.}(2010{\natexlab{a}})\citenamefont{Song,
  Rachel, and {Le Hur}}}]{Song10}
\bibinfo{author}{\bibfnamefont{H.~F.} \bibnamefont{Song}},
  \bibinfo{author}{\bibfnamefont{S.}~\bibnamefont{Rachel}}, \bibnamefont{and}
  \bibinfo{author}{\bibfnamefont{K.}~\bibnamefont{{Le Hur}}},
  \bibinfo{journal}{Phys. Rev. B} \textbf{\bibinfo{volume}{82}},
  \bibinfo{pages}{012405} (\bibinfo{year}{2010}{\natexlab{a}}).

\bibitem[{\citenamefont{Hoyos et~al.}(2011)\citenamefont{Hoyos, Laflorencie,
  Vieira, and Vojta}}]{Hoyos11}
\bibinfo{author}{\bibfnamefont{J.~A.} \bibnamefont{Hoyos}},
  \bibinfo{author}{\bibfnamefont{N.}~\bibnamefont{Laflorencie}},
  \bibinfo{author}{\bibfnamefont{A.~P.} \bibnamefont{Vieira}},
  \bibnamefont{and} \bibinfo{author}{\bibfnamefont{T.}~\bibnamefont{Vojta}},
  \bibinfo{journal}{Euro. Phys. Lett.} \textbf{\bibinfo{volume}{93}},
  \bibinfo{pages}{30004} (\bibinfo{year}{2011}).

\bibitem[{\citenamefont{Klich and Levitov}(2009)}]{Klich09}
\bibinfo{author}{\bibfnamefont{I.}~\bibnamefont{Klich}} \bibnamefont{and}
  \bibinfo{author}{\bibfnamefont{L.}~\bibnamefont{Levitov}},
  \bibinfo{journal}{Phys. Rev. Lett.} \textbf{\bibinfo{volume}{102}},
  \bibinfo{pages}{100502} (\bibinfo{year}{2009}).

\bibitem[{\citenamefont{Song et~al.}(2010{\natexlab{b}})\citenamefont{Song,
  Flindt, Rachel, Klich, and {Le Hur}}}]{Song10-2}
\bibinfo{author}{\bibfnamefont{H.~F.} \bibnamefont{Song}},
  \bibinfo{author}{\bibfnamefont{C.}~\bibnamefont{Flindt}},
  \bibinfo{author}{\bibfnamefont{S.}~\bibnamefont{Rachel}},
  \bibinfo{author}{\bibfnamefont{I.}~\bibnamefont{Klich}}, \bibnamefont{and}
  \bibinfo{author}{\bibfnamefont{K.}~\bibnamefont{{Le Hur}}}
  (\bibinfo{year}{2010}{\natexlab{b}}), \eprint{cond-mat/1008.5191}.

\bibitem[{\citenamefont{Hsu et~al.}(2009{\natexlab{b}})\citenamefont{Hsu,
  Grosfeld, and Fradkin}}]{Hsu09-2}
\bibinfo{author}{\bibfnamefont{B.}~\bibnamefont{Hsu}},
  \bibinfo{author}{\bibfnamefont{E.}~\bibnamefont{Grosfeld}}, \bibnamefont{and}
  \bibinfo{author}{\bibfnamefont{E.}~\bibnamefont{Fradkin}},
  \bibinfo{journal}{Phys. Rev. B} \textbf{\bibinfo{volume}{80}},
  \bibinfo{pages}{235412} (\bibinfo{year}{2009}{\natexlab{b}}).

\bibitem[{\citenamefont{Cardy}(2010)}]{Cardy10}
\bibinfo{author}{\bibfnamefont{J.}~\bibnamefont{Cardy}} (\bibinfo{year}{2010}),
  \eprint{cond-mat/1012.5116}.

\bibitem[{\citenamefont{Furukawa and Kim}(2011)}]{Furukawa11}
\bibinfo{author}{\bibfnamefont{S.}~\bibnamefont{Furukawa}} \bibnamefont{and}
  \bibinfo{author}{\bibfnamefont{Y.~B.} \bibnamefont{Kim}},
  \bibinfo{journal}{Phys. Rev. B} \textbf{\bibinfo{volume}{83}},
  \bibinfo{pages}{085112} (\bibinfo{year}{2011}).

\bibitem[{\citenamefont{Alet et~al.}(2007)\citenamefont{Alet, Capponi,
  Laflorencie, and Mambrini}}]{Alet07}
\bibinfo{author}{\bibfnamefont{F.}~\bibnamefont{Alet}},
  \bibinfo{author}{\bibfnamefont{S.}~\bibnamefont{Capponi}},
  \bibinfo{author}{\bibfnamefont{N.}~\bibnamefont{Laflorencie}},
  \bibnamefont{and} \bibinfo{author}{\bibfnamefont{M.}~\bibnamefont{Mambrini}},
  \bibinfo{journal}{Phys. Rev. Lett.} \textbf{\bibinfo{volume}{99}},
  \bibinfo{pages}{117204} (\bibinfo{year}{2007}).

\bibitem[{\citenamefont{Chhajlany et~al.}(2007)\citenamefont{Chhajlany,
  Tomczak, and W{\'{o}}jcik}}]{Chhajlany07}
\bibinfo{author}{\bibfnamefont{R.~W.} \bibnamefont{Chhajlany}},
  \bibinfo{author}{\bibfnamefont{P.}~\bibnamefont{Tomczak}}, \bibnamefont{and}
  \bibinfo{author}{\bibfnamefont{A.}~\bibnamefont{W{\'{o}}jcik}},
  \bibinfo{journal}{Phys. Rev. Lett.} \textbf{\bibinfo{volume}{99}},
  \bibinfo{pages}{167204} (\bibinfo{year}{2007}).

\bibitem[{\citenamefont{Peschel and Eisler}(2009)}]{Peschel09}
\bibinfo{author}{\bibfnamefont{I.}~\bibnamefont{Peschel}} \bibnamefont{and}
  \bibinfo{author}{\bibfnamefont{V.}~\bibnamefont{Eisler}},
  \bibinfo{journal}{J. Phys. A: Math. Theor.} \textbf{\bibinfo{volume}{42}},
  \bibinfo{pages}{504003} (\bibinfo{year}{2009}).

\bibitem[{\citenamefont{Bombelli et~al.}(1986)\citenamefont{Bombelli, Koul,
  Lee, and Sorkin}}]{Bombelli86}
\bibinfo{author}{\bibfnamefont{L.}~\bibnamefont{Bombelli}},
  \bibinfo{author}{\bibfnamefont{R.~K.} \bibnamefont{Koul}},
  \bibinfo{author}{\bibfnamefont{J.}~\bibnamefont{Lee}}, \bibnamefont{and}
  \bibinfo{author}{\bibfnamefont{R.~D.} \bibnamefont{Sorkin}},
  \bibinfo{journal}{Phys. Rev. D} \textbf{\bibinfo{volume}{34}},
  \bibinfo{pages}{373} (\bibinfo{year}{1986}).

\bibitem[{\citenamefont{Casini and Huerta}(2009)}]{Casini09}
\bibinfo{author}{\bibfnamefont{H.}~\bibnamefont{Casini}} \bibnamefont{and}
  \bibinfo{author}{\bibfnamefont{M.}~\bibnamefont{Huerta}},
  \bibinfo{journal}{J. Phys. A: Math. Theor.} \textbf{\bibinfo{volume}{42}},
  \bibinfo{pages}{504007} (\bibinfo{year}{2009}).

\bibitem[{\citenamefont{Eisert and Cramer}(2005)}]{Eisert05}
\bibinfo{author}{\bibfnamefont{J.}~\bibnamefont{Eisert}} \bibnamefont{and}
  \bibinfo{author}{\bibfnamefont{M.}~\bibnamefont{Cramer}},
  \bibinfo{journal}{Phys. Rev. A} \textbf{\bibinfo{volume}{72}},
  \bibinfo{pages}{042112} (\bibinfo{year}{2005}).

\bibitem[{\citenamefont{Peschel and Zhao}(2005)}]{Peschel05}
\bibinfo{author}{\bibfnamefont{I.}~\bibnamefont{Peschel}} \bibnamefont{and}
  \bibinfo{author}{\bibfnamefont{J.}~\bibnamefont{Zhao}}, \bibinfo{journal}{J.
  Stat. Mech.} \textbf{\bibinfo{volume}{(2005)}}, \bibinfo{pages}{{P11002}}
  (\bibinfo{year}{2005}).

\bibitem[{\citenamefont{O{\'{r}}us et~al.}(2006)\citenamefont{O{\'{r}}us,
  I.Latorre, Eisert, and Cramer}}]{Orus06}
\bibinfo{author}{\bibfnamefont{R.}~\bibnamefont{O{\'{r}}us}},
  \bibinfo{author}{\bibfnamefont{J.}~\bibnamefont{I.Latorre}},
  \bibinfo{author}{\bibfnamefont{J.}~\bibnamefont{Eisert}}, \bibnamefont{and}
  \bibinfo{author}{\bibfnamefont{M.}~\bibnamefont{Cramer}},
  \bibinfo{journal}{Phys. Rev. A} \textbf{\bibinfo{volume}{73}},
  \bibinfo{pages}{060303} (\bibinfo{year}{2006}).

\bibitem[{\citenamefont{Srednicki}(1993)}]{Srednicki93}
\bibinfo{author}{\bibfnamefont{M.}~\bibnamefont{Srednicki}},
  \bibinfo{journal}{Phys. Rev. Lett.} \textbf{\bibinfo{volume}{71}},
  \bibinfo{pages}{666} (\bibinfo{year}{1993}).

\bibitem[{\citenamefont{Barthel et~al.}(2006)\citenamefont{Barthel, Chung, and
  Schollw{\"{o}}ck}}]{Barthel06}
\bibinfo{author}{\bibfnamefont{T.}~\bibnamefont{Barthel}},
  \bibinfo{author}{\bibfnamefont{M.-C.} \bibnamefont{Chung}}, \bibnamefont{and}
  \bibinfo{author}{\bibfnamefont{U.}~\bibnamefont{Schollw{\"{o}}ck}},
  \bibinfo{journal}{Phys. Rev. A} \textbf{\bibinfo{volume}{74}},
  \bibinfo{pages}{022329} (\bibinfo{year}{2006}).

\bibitem[{\citenamefont{White et~al.}(1994)\citenamefont{White, Noack, and
  Scalapino}}]{White94}
\bibinfo{author}{\bibfnamefont{S.~R.} \bibnamefont{White}},
  \bibinfo{author}{\bibfnamefont{R.~M.} \bibnamefont{Noack}}, \bibnamefont{and}
  \bibinfo{author}{\bibfnamefont{D.~J.} \bibnamefont{Scalapino}},
  \bibinfo{journal}{Phys. Rev. Lett.} \textbf{\bibinfo{volume}{73}},
  \bibinfo{pages}{886} (\bibinfo{year}{1994}).

\end{thebibliography}

\end{document}